\documentclass[aps,prapplied,10pt,twocolumn,superscriptaddress,amsmath,amssymb]{revtex4-2}

\usepackage{graphicx}
\usepackage{multirow}
\usepackage[usenames]{color}
\usepackage{ulem}
\usepackage{hyperref}

\usepackage{placeins}

\begin{document}

\title{\texorpdfstring{Superconducting spin valve effect in Fe/Si$_3$N$_4$/Pb/Si$_3$N$_4$/Fe heterostructures}{Superconducting spin valve effect in Fe/Si3N4/Pb/Si3N4/Fe heterostructures}}

\author{A.~A.~Kamashev}
\affiliation{Zavoisky Physical-Technical Institute, FRC Kazan Scientific Center of RAS, 420029 Kazan, Russia}

\author{N.~N.~Garif'yanov}
\affiliation{Zavoisky Physical-Technical Institute, FRC Kazan Scientific Center of RAS, 420029 Kazan, Russia}

\author{A.~A.~Validov}
\affiliation{Zavoisky Physical-Technical Institute, FRC Kazan Scientific Center of RAS, 420029 Kazan, Russia}

\author{A.~S.~Osin}
\affiliation{L.~D.\ Landau Institute for Theoretical Physics RAS, 142432 Chernogolovka, Russia}

\author{Ya.~V.~Fominov}
\affiliation{L.~D.\ Landau Institute for Theoretical Physics RAS, 142432 Chernogolovka, Russia}
\affiliation{Moscow Institute of Physics and Technology, 141701 Dolgoprudny, Russia}
\affiliation{Laboratory for Condensed Matter Physics, HSE University, 101000 Moscow, Russia}

\author{I.~A.~Garifullin}
\affiliation{Zavoisky Physical-Technical Institute, FRC Kazan Scientific Center of RAS, 420029 Kazan, Russia}

\date{4 October 2025}

\begin{abstract}
The structures of the superconducting spin valve (SSV) Fe/\allowbreak Si$_3$N$_4$/\allowbreak Pb/\allowbreak Si$_3$N$_4$/\allowbreak Fe (where Si$_3$N$_4$ is a dielectric insulating layer of controlled thickness) were investigated. The dependence of the magnitude of the SSV effect  on the thicknesses of the superconducting (S) and insulating (I) layers was studied. Optimization of the S and I layer thicknesses enabled a complete switching between the normal and superconducting states when the mutual orientation of the magnetizations of the ferromagnetic (F) layers changed from antiparallel to parallel. A maximal SSV effect value of 0.36\,K  was achieved in an external magnetic field of 1\,kOe. These results demonstrate that SSV structures with tunable S/F interface transparency controlled by insulating interlayers are promising for achieving a significant magnitude of the effect. This opens new avenues for the development of such systems and their potential applications in spintronic devices.
\end{abstract}

\maketitle

\section{Introduction}

The superconducting spin valve (SSV) effect arises from different degrees of suppression of superconductivity by ferromagnetism depending on whether the mutual orientation of the magnetizations of the ferromagnetic (F) layers is parallel (P) or antiparallel (AP) in F1/F2/S or F1/S/F2 thin-film heterostructures \cite{Oh,Tagirov,Buzdin2}. This effect results in different superconducting transition temperatures ($T_c$) for P ($T_c^\mathrm{P}$) and AP ($T_c^\mathrm{AP}$) orientations of magnetizations of ferromagnetic layers in SSV structures. The operation principle of SSV structures is based on the superconductor/ferromagnet (S/F) proximity effect. A variety of systems can be designed using the S/F proximity effect, which are considered as fundamental building blocks for superconducting spintronic devices (see, e.g., Refs.\ \cite{Ioffe,Feigelman,Buzdin1,Linder,Eschrig,Ryazanov1999,Ryazanov2001a,Gu2002,Zdravkov2013,Lenk2017,Gaifullin,Kamashev20191,Kamashev20241}). For instance, a superconducting spin valve shows promise as an element in superconducting spintronics, such as a superconducting switcher or transistor. 

The magnitude of the ordinary SSV effect ($\Delta T_c$) is defined as the difference between $T_c^\mathrm{P}$ and $T_c^\mathrm{AP}$. If  the condition $\Delta T_c > \delta T_c$ is satisfied (where $\delta T_c$ is the width of the superconducting transition), it becomes possible to switch the superconducting state on/off in SSV structures by changing the mutual orientation of the magnetizations from P to AP (referred to as the full SSV effect). A wide range of heterostructures based on S/F interfaces with direct and high-quality contact between the S and F layers has been extensively studied  (see, e.g., Refs.\ \cite{Kamashev20191,Gu2002,Eschrig2,Moraru,Steiner,Stoutimore,Pugach2009,Leksin2010,Fominov,Leksin2011,Kamashev2019,Gu2015,Aarts2015,Nano,Halterman4,Halterman5}).

While the proximity effect in structures with metallic ferromagnets leads to the SSV effect, it can also have a detrimental influence on a system's superconductivity in general. Both $T_c^\mathrm{P}$ and $T_c^\mathrm{AP}$ can be strongly suppressed compared with the critical temperature of the S material.
From this point of view, SSV structures of the FI/S/FI type (where FI is a ferromagnetic insulator) may seem advantageous \cite{DeGennes1966,Li2013,Zhu2017}. In such structures, the direct proximity effect is absent while $T_c$ is controlled by the relative orientation of the interfacial exchange fields (felt by electrons upon reflecting from the FI layers).
Alternatively, the overall suppression of $T_c$ can be mitigated due to introducing nonmagnetic insulating layers (I) at the S/F interfaces with metallic ferromagnets. Indeed, the resulting F/I/S/I/F structures demonstrate significant SSV effect \cite{Deutscher,Miao2007,Luo2010,Kamashev20242,Kamashev2024}. 
At the same time, the F/I/S/I/F configuration offers a crucial benefit: by optimizing parameters, in particular, the thickness of the S layer, superconductivity can be made highly sensitive to the magnetic state of the ferromagnets. The inclusion of the I layers then provides a means to tune such ``fragile'' superconductivity in F/I/S/I/F structures and, therefore, to reach significant values of $\Delta T_c$.

A pronounced SSV effect in such F/I/S/I/F structures was first demonstrated in Ref.\ \cite{Deutscher}. Although similar structures were subsequently investigated  \cite{Miao2007,Luo2010}, the reported values of $\Delta T_c$ turned out to be much smaller than in Ref.\ \cite{Deutscher}, and the full SSV effect was not reproduced. Important confirmations of the findings of Ref.\ \cite{Deutscher}, including comparable values of $\Delta T_c$ and the full SSV effect, were recently achieved in Refs.\ \cite{Kamashev20242,Kamashev2024}.

However, in Refs.\ \cite{Deutscher,Kamashev20242,Kamashev2024}, significant SSV effects were observed in structures with oxidized S/F interfaces. While these results have been consistently reproduced in SSV structures with different combinations of ferromagnetic and superconducting materials, the properties of the resulting insulating layers (in particular, their thicknesses) were not precisely controlled, due to the specific preparation methods used.

To address this issue, in the present work, the superconducting and ferromagnetic layers were intentionally separated by  insulating layers of a well-controlled thickness at the S/F interfaces of the SSV structures. Model SSV structures of  Fe/\allowbreak Si$_3$N$_4$/\allowbreak Pb/\allowbreak Si$_3$N$_4$/\allowbreak Fe (where Si$_3$N$_4$ is the dielectric insulating layers) with varying thicknesses of both the superconducting and insulating layers were fabricated and the properties of the Fe/\allowbreak Si$_3$N$_4$/\allowbreak Pb/\allowbreak Si$_3$N$_4$/\allowbreak Fe SSV structures have been investigated. 
(Note that previously, Si$_3$N$_4$ in SSV structures mainly served as a protective layer, not being utilized as a functional insulating material.)
The dependence of the magnitude $\Delta T_c$ of the SSV effect on the thickness of Si$_3$N$_4$ and Pb layers has been studied. It was found that the value of $\Delta T_c$ reaches up to 0.36\,K for optimal parameters of the SSV structure yielding the full SSV effect in the external magnetic field of 1\,kOe. Such values of the effect exceed the values previously reported in S/I/F/I/S structures \cite{Deutscher,Miao2007,Luo2010,Kamashev20242,Kamashev2024} and also exceed most of the values of the ordinary SSV effect in the structures with a direct high-quality contact between S and F layers.

\section{Samples}

\begin{figure}[t]
\includegraphics[width=0.5\columnwidth]{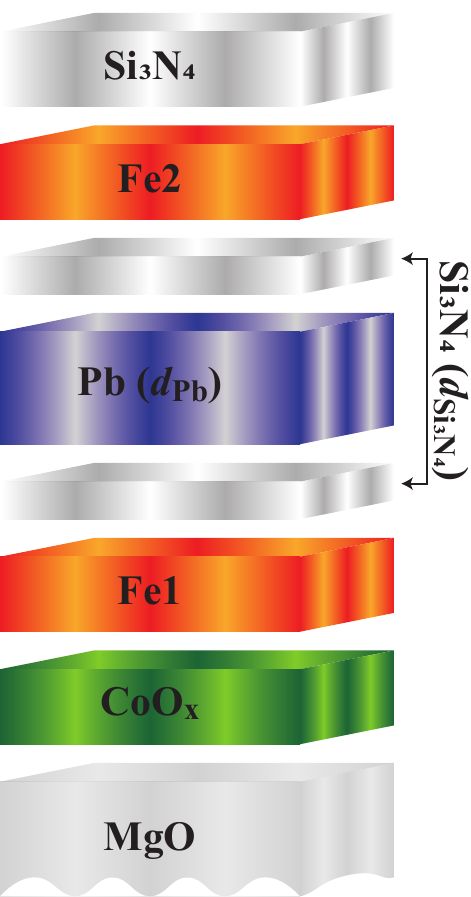}
\caption{Design of the prepared SSV heterostructures: CoO$_x$(3.5nm)/\allowbreak Fe1(3nm)/\allowbreak Si$_3$N$_4$1($d_\mathrm{Si_{3}N_{4}}$)/\allowbreak Pb($d_\mathrm{Pb}$)/\allowbreak Si$_3$N$_4$2($d_\mathrm{Si_{3}N_{4}}$)/\allowbreak Fe2(3nm)/\allowbreak Si$_3$N$_4$(85nm) with variable Si$_3$N$_4$ layers thickness $d_\mathrm{Si_{3}N_{4}}$ in the range from 0 to 1.2\,nm and Pb layer $d_\mathrm{Pb}$ thickness in the range of 40 to 60\,nm.}
\label{fig1}
\end{figure}

The samples were produced using the  deposition system from BESTEC GmbH at Zavoisky Physical-Technical Institute. Its construction allows us to transfer samples between three chambers during one vacuum cycle: the load lock station with vacuum shutters and vacuum about $1\times 10^{-8}$\,mbar, the molecular-beam epitaxy (MBE) chamber with the stationary vacuum of about $1\times10^{-10}$\,mbar, and the magnetron sputtering (dc and ac) chamber with the stationary vacuum of about $1\times 10^{-9}$\,mbar. The prepared heterostructures consist of three series and have the following composition: CoO$_x$(3.5nm)/\allowbreak Fe1(3nm)/\allowbreak Si$_3$N$_4$1($d_\mathrm{Si_{3}N_{4}}$)/\allowbreak Pb($d_\mathrm{Pb}$)/\allowbreak Si$_3$N$_4$2($d_\mathrm{Si_{3}N_{4}}$)/\allowbreak Fe2(3nm)/\allowbreak Si$_3$N$_4$(85nm) with the variable Si$_3$N$_4$ layers thickness $d_\mathrm{Si_{3}N_{4}}$ in the range from 0 to 1.2\,nm and the Pb layer  thickness $d_\mathrm{Pb}$ in the range of 40 to 60\,nm. The heterostructures were deposited on high-quality single-crystalline MgO(001) substrates. All materials used for evaporation had a purity of better than 4N (99.99\,at.\,\%). The general design of all series of the samples is presented in Fig.~\ref{fig1}. There,  CoO$_x$ is an antiferromagnetic layer necessary for fixing the direction of magnetization of the Fe1 layer, Fe1 and Fe2 are the ferromagnetic F1 and F2 layers, Si$_3$N$_4$1 and Si$_3$N$_4$2 are the  dielectric insulating interlayers, Pb is the superconducting S layer, and the top  Si$_3$N$_4$ layer is the protective layer. The layers were deposited by e-beam evaporation (Co, Fe, Pb) and ac magnetron sputtering (Si$_3$N$_4$) methods. 

The sequence of the sample preparation was as follows. The first layer of CoO$_x$ was prepared in two steps: (a)~at first, a metallic Co layer was deposited in an MBE chamber; (b)~after that, the sample was oxidized in an oxygen atmosphere at 100\,mbar for two hours in a load lock station. Next, subsequent layers of SSV structures were fabricated. The Pb layer and the subsequent layers of the structures were deposited at a reduced substrate temperature of $T_\mathrm{sub}\approx 150$\,K. This procedure is necessary to form the smoothest Pb layer according to Ref.\ \cite{Nano}. The deposition rates were as follows: 0.5\,{\AA}/s for the Co, Fe, Si$_3$N$_4$1, and Si$_3$N$_4$2 layers; 12\,{\AA}/s for the Pb layer; 2\,{\AA}/s for the Si$_3$N$_4$ protective layer.

The residual resistivity ratio $\mathrm{RRR} =R_{300\mathrm{K}}/R_{10\mathrm{K}}$ of the investigated samples was found to be in the range of 15--20. Such values indicate high purity and high quality of the Pb layer in the prepared samples.

Table \ref{ParametersTab} summarizes the parameters of all series of samples.

\begin{table}[t]
\begin{tabular}{|c|c|c|c|c|c|}
\hline
                     \begin{tabular}[c]{@{}c@{}}Series \\ number \end{tabular} &
                     \begin{tabular}[c]{@{}c@{}}Sample \\ number \end{tabular} & 
                     \begin{tabular}[c]{@{}c@{}}$d_{\mathrm{Pb}}$\\(nm)\end{tabular} & 
                     \begin{tabular}[c]{@{}c@{}}$d_\mathrm{Si_3N_4}$\\(nm)\end{tabular} & 
                     \begin{tabular}[c]{@{}c@{}}$\Delta T_c$\\(K)\end{tabular} & 
                     \begin{tabular}[c]{@{}c@{}}$T_c$ (K)\\($H_0 = 0$\,Oe)\end{tabular} \\ \hline
                    
\multirow{4}{*}{1} & 1 & \multirow{4}{*}{40} & 0     &  0    & $\leqslant$ 1.4  \\ \cline{2-2} \cline{4-6} 
                   & 2 &                     & 0.3   & 0.21  & 5.4         \\ \cline{2-2} \cline{4-6} 
                   & 3 &                     & 0.6   & 0.08  & 6.7         \\ \cline{2-2} \cline{4-6} 
                   & 4 &                     & 1.2   &  0    & 7           \\ \hline

\multirow{5}{*}{2} & 1 & \multirow{5}{*}{50} & 0     &  0    & $\leqslant$ 1.4  \\ \cline{2-2} \cline{4-6} 
                   & 2 & & 0.15                &  0    &        $\leqslant$ 1.4  \\ \cline{2-2} \cline{4-6} 
                   & 3 & & 0.3                 & 0.36  & 6.08               \\ \cline{2-2} \cline{4-6} 
                   & 4 & & 0.6                 &  0.1  & 6.76               \\ \cline{2-2} \cline{4-6} 
                   & 5 & & 1.2                 &  0    & 7                  \\ \hline

\multirow{4}{*}{3} & 1 & \multirow{4}{*}{60} & 0     &  0.090    & 3.3         \\ \cline{2-2} \cline{4-6} 
                   & 2 & & 0.3                 & 0.078 & 6.53               \\ \cline{2-2} \cline{4-6} 
                   & 3 & & 0.6                 & 0.043 & 6.83               \\ \cline{2-2} \cline{4-6}  
                   & 4 & & 1.2                 & 0     & 7                  \\ \hline 
  
\end{tabular}
\caption{Parameters of the prepared SSV heterostructures CoO$_x$(3.5nm)/\allowbreak Fe1(3nm)/\allowbreak Si$_3$N$_4$1($d_\mathrm{Si_{3}N_{4}}$)/\allowbreak Pb($d_\mathrm{Pb}$)/\allowbreak Si$_3$N$_4$2($d_\mathrm{Si_{3}N_{4}}$)/\allowbreak Fe2(3nm)/\allowbreak Si$_3$N$_4$(85nm) with variable Si$_3$N$_4$ layers thickness $d_\mathrm{Si_{3}N_{4}}$ in the range from 0 to 1.2\,nm and Pb layer $d_\mathrm{Pb}$ thickness in the range of 40 to 60\,nm.} 
\label{ParametersTab} 
\end{table}

\section{Experimental results}

\begin{figure}[t]
\center{\includegraphics[width=0.97\columnwidth]{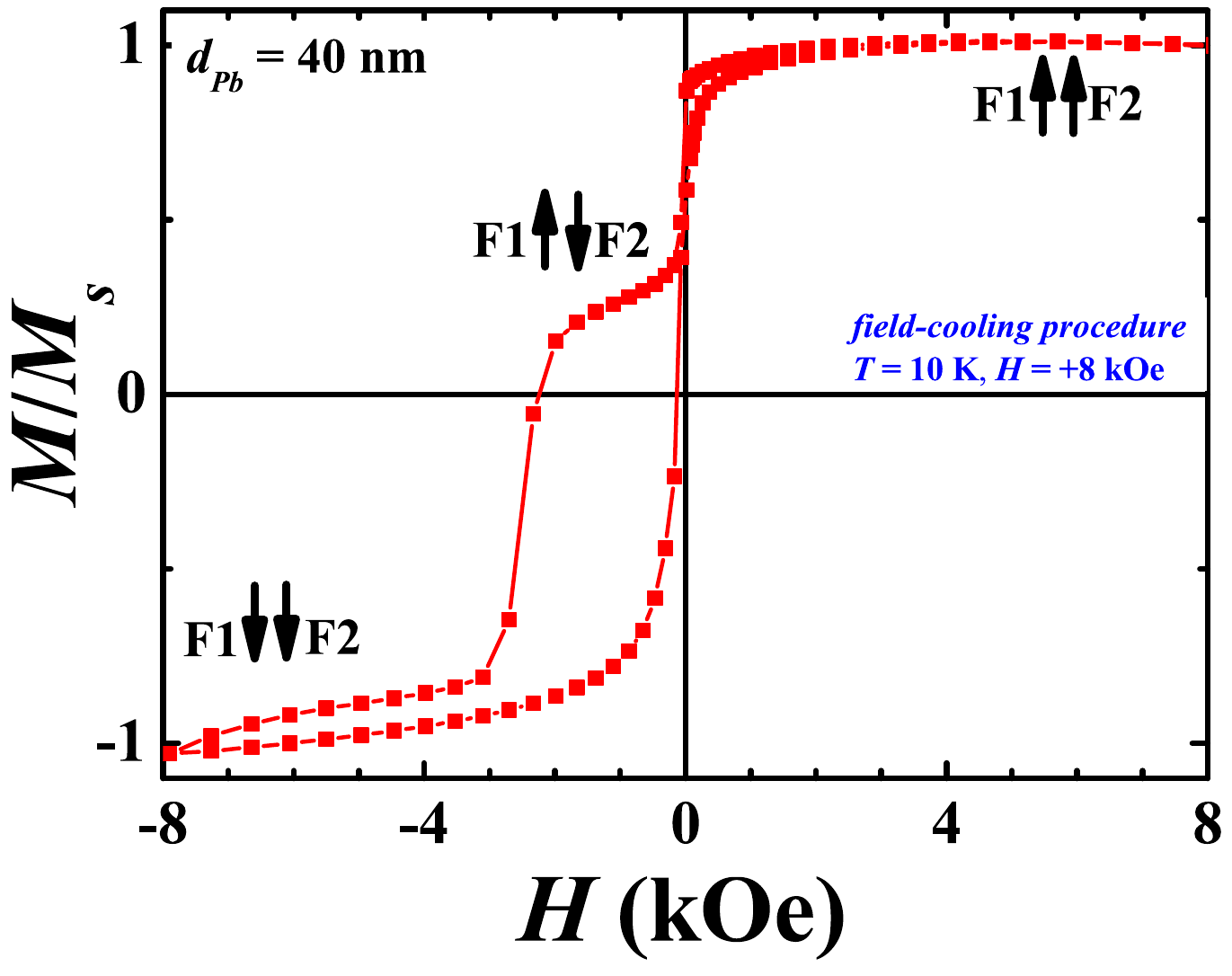}}
\caption{Magnetic hysteresis loop $M(H)$ for the sample Series~1-3 measured after the field cooling procedure from room temperature to $T = 10$\,K in a magnetic field of $+8$\,kOe.}
\label{fig2}
\end{figure}

The study of the magnetic properties of the samples was carried out using a commercial VSM SQUID magnetometer. For all samples, the dependence of the magnetization $M$ on the magnetic field $H$ was measured to determine the range of operational external magnetic fields, i.e., the fields at which the P and AP orientations of the F layers magnetizations are achieved. As an example, Fig.~\ref{fig2} shows the magnetic hysteresis loop for the sample Series~1-3 (see  Table~\ref{ParametersTab}). This type of magnetic hysteresis loops is typical for all series of samples. The samples were cooled from room temperature to 10\,K in a magnetic field of 8\,kOe (field cooling procedure) before measuring the magnetic properties. This procedure is necessary to align the magnetizations of the F layers parallel to each other and to fix the magnetization of the Fe1 layer via exchange bias due to the adjacent antiferromagnetic CoO$_x$ layer (with N\'eel temperature $T_\mathrm{N}^{\mathrm{CoO}_x} \approx 250$\,K). The $M(H)$ dependence was measured at a temperature of 10\,K, with the magnetic field varied from $+8$ to $-8$\,kOe and back to $+8$\,kOe. Both field values of $+8$ and $-8$\,kOe correspond to the P orientation of the magnetizations of the F layers. As can be seen in Fig.~\ref{fig2},  the total magnetization of the sample $M(H)$ is maximum and  constant when the field decreases from $+8$ to $+0.5$\,kOe evincing that the magnetizations of the F layers are parallel and saturated. When the magnetic field is further reduced, $M(H)$ starts to decrease, which corresponds to the continuous reversal of the magnetization of the ``free'' F2 layer, while the magnetization of the F1 layer pinned by the CoO$_x$ layer  remains fixed in the original direction until a magnetic field of $-1.7$\,kOe is applied. Such AP orientation of the F layers' magnetizations is maintained in the range of magnetic fields from $-0.15$ to $-1.7$\,kOe. The magnetization of the F1 layer begins to rotate in the field direction  with  increasing   the field from $-1.7$ to $-3$\,kOe yielding further reduction of $M(H)$ toward negative values. Finally,  in the range of magnetic fields from $-3$ to $-8$\,kOe, the P orientation of the magnetizations of the F layers in the direction opposite to the initial P orientation is realized corresponding to the maximum negative value of $M(H)$.

The superconducting properties of the samples were investigated using the standard four-point method with dc current. The current and voltage leads were attached to the samples using clamping contacts. A highly homogeneous vector electromagnet from Bruker Instruments was employed for measurements in an external magnetic field. The critical temperature $T_c$ was determined by evaluating the resistance as a function of temperature. The $T_c$ value was defined as the midpoint of the superconducting transition. The samples were cooled from room  to helium temperature in an external magnetic field of approximately 5\,kOe, applied along the sample plane. This procedure was similar to that used for magnetic measurements.

\begin{figure}[t]
\center{\includegraphics[width=1\columnwidth]{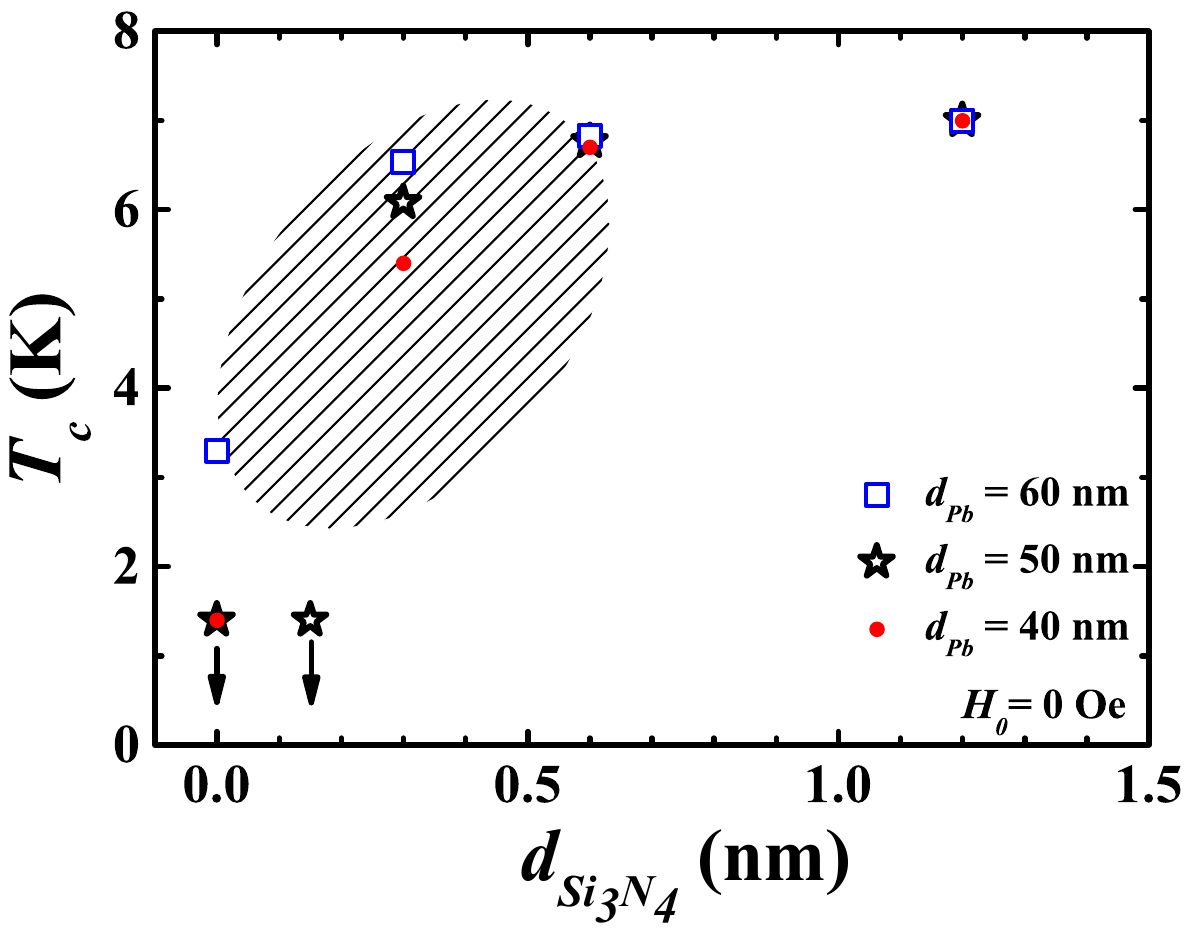}}
\caption{Dependence of $T_c$ on the thickness $d_\mathrm{Si_{3}N_{4}}$ for three series of samples measured without external magnetic field ($H_0 = 0$\,Oe). Shaded area is the optimal range of insulating layer thicknesses for observing the SSV effects (see the text for details).}
\label{fig3}
\end{figure}

Figure~\ref{fig3} shows the dependence of $T_c$ on the thickness of the Si$_3$N$_4$ layer $d_\mathrm{Si_{3}N_{4}}$ for three series of samples measured in the absence of an external magnetic field ($H_0 = 0$\,Oe). According to the figure, for structures with large $d_\mathrm{Si_{3}N_{4}}$ thicknesses, the $T_c$ value is close to that of the bulk lead (7.2\,K). As $d_\mathrm{Si_{3}N_{4}}$ decreases, the $T_c$ value also decreases. For small or zero values of $d_\mathrm{Si_{3}N_{4}}$, superconductivity could not be detected down to the lowest attainable temperature of 1.4\,K.

\begin{figure}[t]
\center{\includegraphics[width=1\columnwidth]{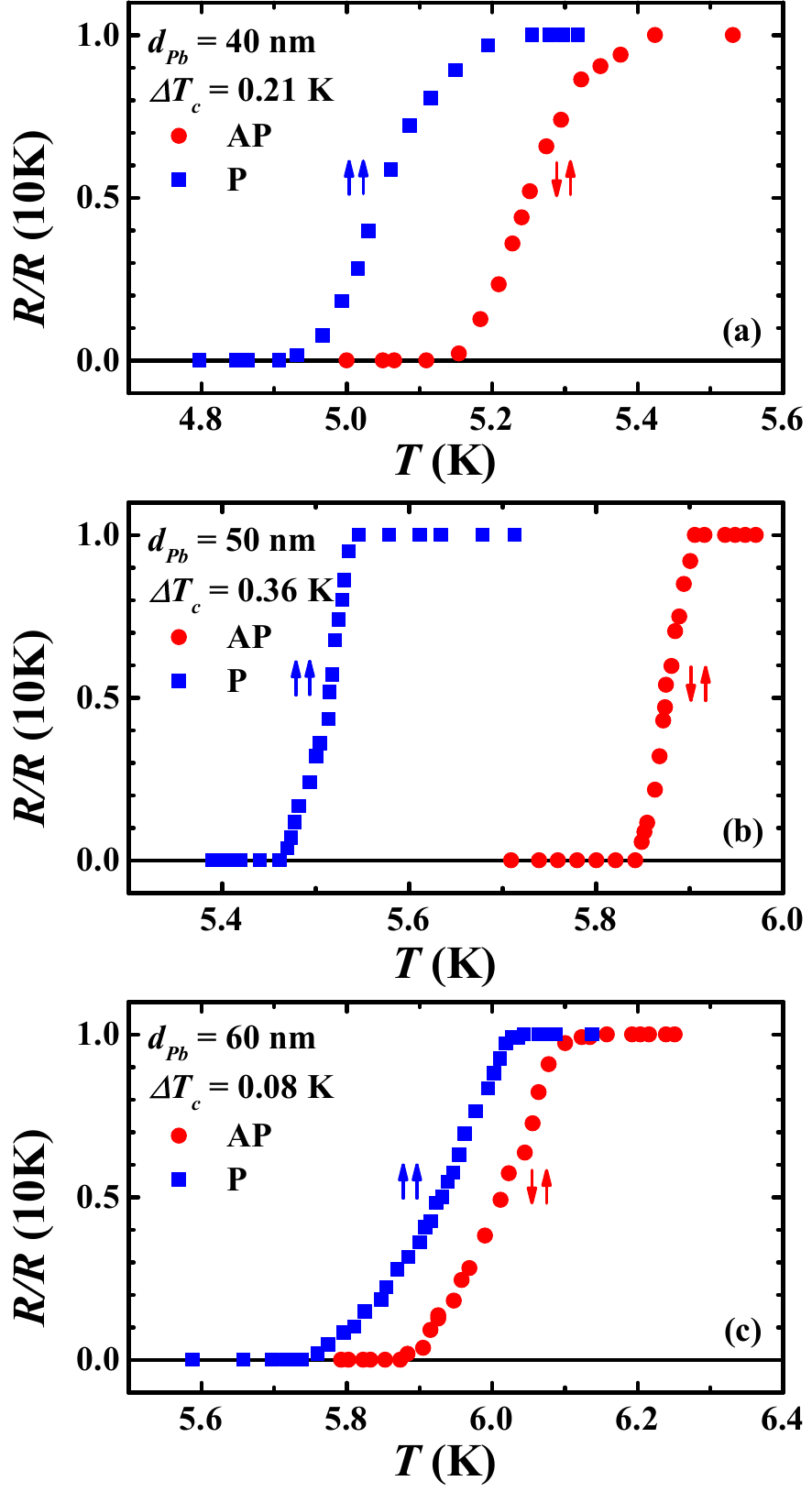}}
\caption{Superconducting transitions curves for the P and AP orientations of the Fe1 and Fe2 layers magnetization for three samples (a)~Series~3-2 ($d_\mathrm{Pb} = 60$\,nm), (b)~Series~2-3 ($d_\mathrm{Pb} = 50$\,nm), and (c)~Series~1-2 ($d_\mathrm{Pb} = 40$\,nm).}
\label{fig4}
\end{figure}

Figure~\ref{fig4} presents the superconducting transitions for the P  and AP orientations of the magnetizations of the Fe1 and Fe2 layers for three samples: (a)~Series~3-2 ($d_\mathrm{Pb} = 60$\,nm), (b)~Series~2-3 ($d_\mathrm{Pb} = 50$\,nm), and (c)~Series~1-2 ($d_\mathrm{Pb} = 40$\,nm). Variation of the mutual orientation of the magnetizations of the Fe layers was achieved by the in-plane rotation of the vector of the applied external magnetic field, $H_0 = 1$\,kOe, which, in turn, caused a corresponding rotation of the Fe2 layer magnetization from the AP to P orientation with respect to the pinned magnetization of the Fe1 layer. As seen in Fig.~\ref{fig4}, the full ordinary SSV effect  ($\Delta T_c \geqslant \delta T_c$) is observed for the sample Series~2-3, with a value of $\Delta T_c \approx 0.36$\,K. 

\begin{figure}[t]
\center{\includegraphics[width=\columnwidth]{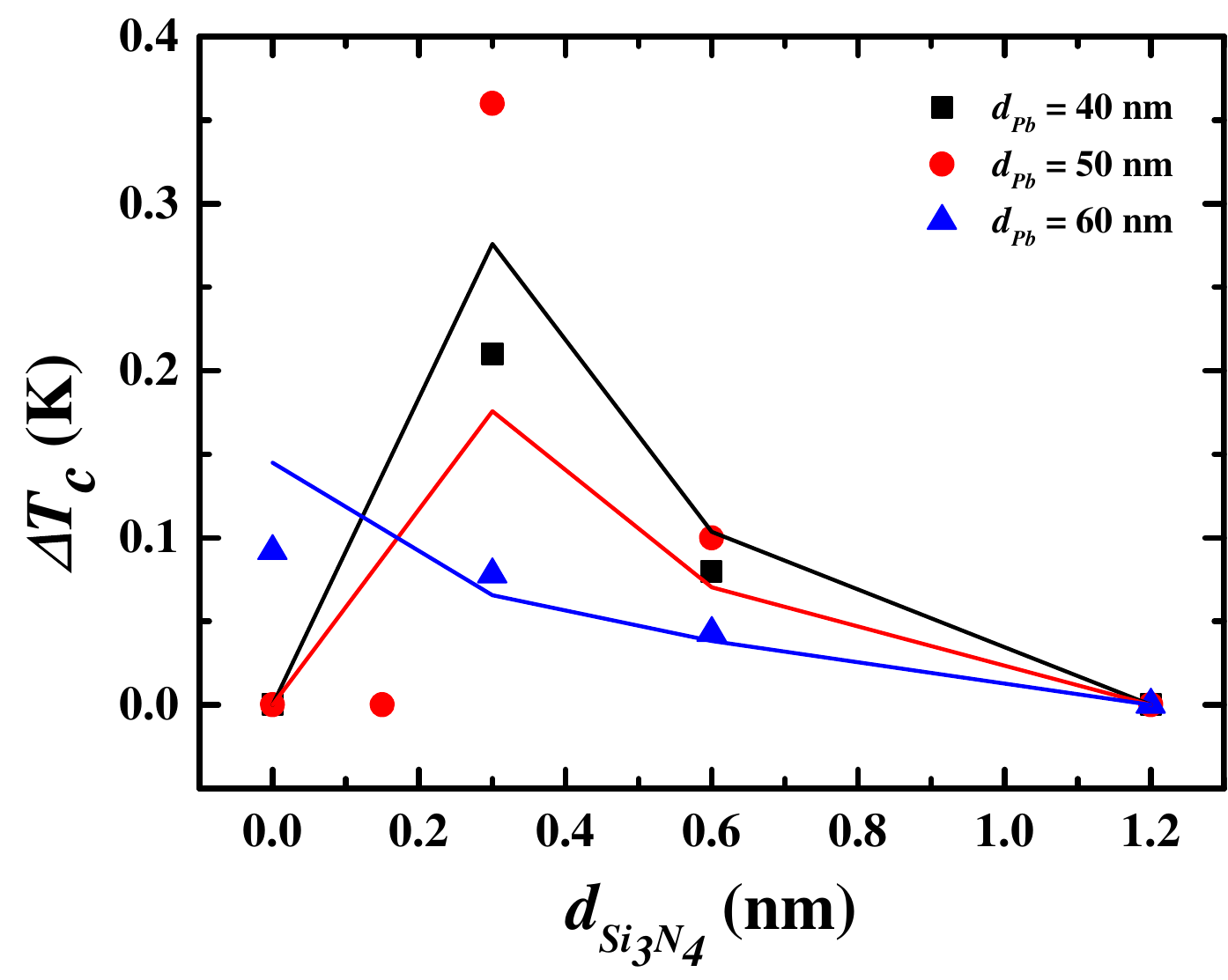}}
\caption{Dependence of the magnitude  of the SSV effect $\Delta T_c$ on the thickness of the Si$_3$N$_4$ layers $d_\mathrm{Si_{3}N_{4}}$ for three series of samples. 
The calculated theoretical points correspond only to $d_{\text{Si}_{3}\text{N}_{4}} =0$, $0.3$, $0.6$, and $1.2$\,nm; they are connected by straight solid lines to aid visual comparison with the experimental data. The fitting parameters are discussed in the main text.}
\label{fig5}
\end{figure}

Figure~\ref{fig5} shows the dependence of $\Delta T_c$ on $d_\mathrm{Si_{3}N_{4}}$ for three series of samples. The maximum value of the SSV effect $\Delta T_c = 0.36$\,K is achieved at the S layer thickness $d_\mathrm{Pb} = 50$\,nm and the I layer thickness $d_\mathrm{Si_{3}N_{4}} = 0.3$\,nm (sample Series~2-3).

Experimental data are available upon request.

\section{Discussion}\label{Discussion}

Figure~\ref{fig3}  demonstrates that for large thicknesses of the Si$_3$N$_4$ layers $d_\mathrm{Si_{3}N_{4}}$, the S/F proximity effect is practically negligible, resulting in the relatively high values of $T_c$ (approximately 7.2\,K). As the thickness of the Si$_3$N$_4$ layer decreases, the S/F proximity effect becomes more pronounced, leading to the suppression of $T_c$ and the inability to observe superconductivity for small (or zero) values of $d_\mathrm{Si_{3}N_{4}}$. This behavior of $T_c$ is characteristic across all sample series and aligns well with the modern understanding of the S/F proximity effect. The closer the superconducting and ferromagnetic layers are to each other, the more pronounced the S/F proximity effect becomes. It is also evident from Fig.~\ref{fig3} that as the thickness of the S layer $d_\mathrm{Pb}$ decreases, the critical temperature $T_c$ decreases as well. This behavior is expected since thinner superconducting layers are more fragile and sensitive to the magnetic component of the system. Thus, based on the dependence shown in Fig.~\ref{fig3}, the optimal range of insulating layer thicknesses for observing the SSV effects can be determined. This range is marked in Fig.~\ref{fig3} as the shaded area. From the dependence shown in Fig.~\ref{fig3}, we can also conclude that we precisely control the parameters of our structures, in particular the thicknesses of the S and I layers. 

Figure~\ref{fig5} shows the dependence of the SSV effect $\Delta T_c$ on the thickness of the Si$_3$N$_4$ layers $d_\mathrm{Si_{3}N_{4}}$ for three series of samples. Significant SSV effects were observed for all series. The absence of the SSV effect in samples with an increased thickness of the I layer is attributed to the suppression of the S/F proximity effect. This suppression arises due to the enhanced spatial separation between S and F layers, which significantly weakens their interaction, as evinced by the data presented in Fig.~\ref{fig3}. 

Similarly, the SSV effect was not observed in samples with ultrathin I layers, where the excessively strong S/F proximity effect leads to a near-complete suppression of superconductivity, as illustrated in Fig.~\ref{fig3}. A nonmonotonic dependence of SSV effect on the Si$_3$N$_4$ layer thickness is evident across all three sample series (see Fig.~\ref{fig5}). The SSV effect is absent at larger $d_\mathrm{Si_{3}N_{4}}$, becomes progressively more pronounced with decreasing thickness, reaches an optimal value at an intermediate $d_\mathrm{Si_{3}N_{4}}$, and subsequently diminishes for minimal or vanishing  $d_\mathrm{Si_{3}N_{4}}$. This behavior underlines the critical role of layer thickness in balancing proximity effects and sustaining superconductivity necessary for the SSV phenomenon. The largest SSV effects were observed for samples from Series~2, which is consistent with the marked area in Fig.~\ref{fig3}. The maximum value of the SSV effect, $\Delta T_c = 0.36$\,K, was recorded for the Series~2-3 sample. The full SSV effect was also observed for this sample [see Fig.~\ref{fig4}(b)]. 

It is noteworthy that the maximum SSV effects occur in samples with an insulating layer thickness $d_\mathrm{Si_{3}N_{4}}$ of approximately 0.3\,nm. At this scale, approaching the atomic level, the layer may exhibit nonuniformity or nonstoichiometry properties. While a comprehensive investigation of these potential structural characteristics is beyond the scope of this work and deserves future study, we note that our prior transmission electron microscopy studies of analogous SSV structures with similar materials demonstrated the absence of significant interdiffusion between layers \cite{Nano}. In the present work, the layer thicknesses were controlled via deposition time, calibrated to a known growth rate. Therefore, the reported thicknesses represent nominal values based on this calibrated process. At the same time, we cannot rule out the presence of weak places in very thin I layers. Still, as a result of the deposition procedure, the insulation between the S and F layers increases, and this effect monotonically grows with increasing the deposition time (i.e., the nominal thickness of the I layers).

An accurate theoretical description of the SSV effect in the studied structures
presents significant challenges due to the complex nature of the interfaces.
While the formation of insulating boundaries is experimentally feasible,
their key characteristics (such as thickness, uniformity, composition) and
physical properties cannot be precisely controlled at present.
Therefore, we focus on analyzing how well a theoretical model can qualitatively
describe the main dependencies of $T_{c}$ and $\Delta T_{c}$ on
the thickness of the insulating layers. 

The theory of the SSV effect for symmetric
F1/S/F2 structures was developed in Refs.\ \cite{Tagirov,Buzdin2,Baladie2003,Fominov2003}. In our analysis below, we follow the method of Ref.\ \cite{Fominov2003}, while increasing I layers' thickness $d_\mathrm{Si_{3}N_{4}}$ is modeled as increasing interface resistance (which enters the boundary conditions). Taking into account a possible inhomogeneity of the interfaces (discussed above), we should consider the interface resistances in our model as an effective global characteristic (reflecting local point-dependent interface resistivities in an averaged manner).

\begin{figure}[t]
\center{\includegraphics[width=1\columnwidth]{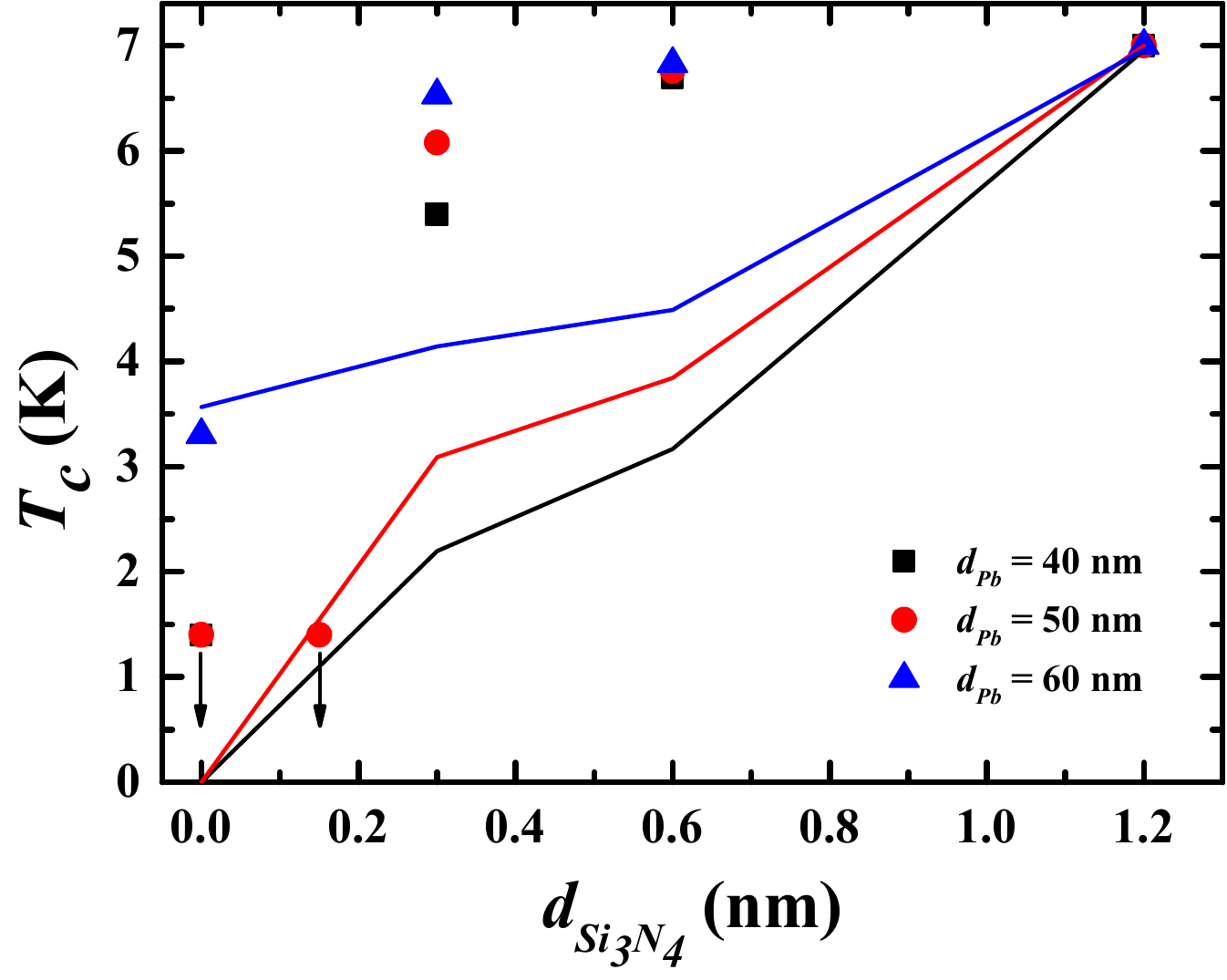}}
\caption{$T_c$ in the P configuration as a function of the interface thickness. The calculated theoretical points correspond only to $d_{\text{Si}_{3}\text{N}_{4}} =0$, $0.3$, $0.6$, and $1.2$\,nm; they are connected by straight solid lines to aid visual comparison with the experimental data. The fitting parameters are discussed in the main text. With the fitting parameters, mainly aimed at explaining the $\Delta T_{c}(d_\mathrm{Si_{3}N_{4}})$ dependence (see Fig.~\ref{fig5}), theory correctly captures the main trends of $T_{c}(d_\mathrm{Si_{3}N_{4}})$ observed in experiment.}
\label{TcFig}
\end{figure}

The theoretical
dependencies $T_{c}(d_\mathrm{Si_{3}N_{4}})$ and $\Delta T_{c}(d_\mathrm{Si_{3}N_{4}})$
obtained from this model exhibit qualitative agreement with the experimental
results, see Figs.~\ref{fig5} and~\ref{TcFig}. The main trends observed in the plots can be summarized as follows. Starting from a state with suppressed superconductivity at $d_\mathrm{Si_{3}N_{4}}=0$\,nm and increasing this parameter, we find initial enhancement of the SSV effect reaching a peak at a certain point: in the P configuration, superconductivity remains suppressed, while in the AP configuration, it is already present. A further increase of the interface thickness weakens the superconducting proximity effect, causing $T_c$ to approach its bulk value, while the SSV effect $\Delta T_c$ is suppressed due to reduced sensitivity to the relative orientation of the F layers. The theory thus explains nontrivial nonmonotonic dependence of the SSV effect on the interface resistance, i.e., the $\Delta T_{c}(d_\mathrm{Si_{3}N_{4}})$ dependence.

Notably, for $T_{c}$, the model correctly predicts the order of magnitude and qualitative behavior of the effect, see Fig.~\ref{TcFig}. At the same time,
for $\Delta T_{c}$, even a satisfactory quantitative agreement is achieved
for samples from Series~1 and~3, see Fig.~\ref{fig5}. The model's fitted parameters are: $\gamma=0.038$, $h=0.075\,\text{eV}$, $\xi_{S}=40\,\text{nm}$,
$\xi_{F}=5\,\text{nm}$, and the values of the interface-resistance parameter $\gamma_{b}$ are given
in Table \ref{GammaTab}. The precise definitions of the fitting parameters and details of the fitting procedure are given in Appendix.

As demonstrated by Fig.~\ref{fig5}, the most pronounced SSV effect is observed for the Series~2 samples. However, for this very series, the theoretical fit shows the largest deviation from the experimental points. Such a deviation is a consequence of our stringent fitting procedure, specifically, the constraints imposed on the possible values of $\gamma_b$. According to this procedure, we take care of correlations between the $\gamma_b$ values for the three series (see Appendix for details). At the same time, if we allow ourselves to fit $\gamma_b$ within each series independently (taking into account possible uncertainties of the interface quality and homogeneity in the case of very thin insulating layers), then we can achieve good quantitative agreement for all the three series (including Series~2), as shown in Fig.~\ref{fig:DeltaTc_app_fit}(a) in Appendix.

\begin{table}[t] \centering
\begin{tabular}{|c|c|c|c|c|} 
\hline 
\textbf{$d_{\text{Si}_{3}\text{N}_{4}}=$} & \textbf{0\,\text{nm}} & \textbf{0.3\,\text{nm}} & \textbf{0.6\,\text{nm}} & \textbf{1.2\,\text{nm}} \\ 
\hline 
\textbf{Series 1} &0  &0.17  &0.23  &6.0  \\ 
\hline 
\textbf{Series 2} &0  &0.14  &0.2  &6.0  \\ 
\hline 
\textbf{Series 3} &0.11  &0.16  &0.2  &4.0  \\ 
\hline 
\end{tabular}
\caption{Fitted values of the interface-resistance parameter $\gamma_b$  for samples from different series.} 
\label{GammaTab} 
\end{table}

\section{Conclusions}

The magnetic and superconducting properties of SSV heterostructures Fe/\allowbreak Si$_3$N$_4$/\allowbreak Pb/\allowbreak Si$_3$N$_4$/\allowbreak Fe, where Si$_3$N$_4$ are dielectric insulating layers of controlled thickness, were investigated. The dependence of the value of the SSV effect on the thicknesses of the superconducting and insulating layers was studied. Significant values of the SSV effect were observed in all prepared series of samples. The full SSV effect with the value $\Delta T_c = 0.36$\,K was recorded for the sample with the optimal parameters of the S and I layers. 

The present study builds upon our previous research, which explored the SSV effects in structures with oxidized S/F interfaces \cite{Kamashev20242,Kamashev2024}. In those studies \cite{Kamashev20242,Kamashev2024}, control over the parameters of the oxidized interlayers was limited due to technological constraints of sample preparation. In contrast, in the present work, it was possible to manage the parameters of the S/F interface in our structures. 

The observation of significant SSV effect values and its features in structures with various types of barriers at the S/F interfaces is qualitatively explained by the proximity-effect-based theory, while quantitative agreement is only partial. This can be due to both simplifying assumptions of  theory and complicated nature of actual interfaces in experiment.

It is worth noting that in our SSV structures, large effect magnitudes are achieved in relatively small magnetic fields. Additionally, such pronounced SSV effects are observed in structures employing conventional elemental ferromagnetic and superconducting materials, which may simplify the fabrication process for efficient SSV structures in future applications.

\acknowledgments

The authors would like to thank V.\ Kataev for useful discussions.
The work of A.A.K.\ concerning the preparation of the samples was funded by the Russian Science
Foundation according to Research Project No.\ 25-72-10025. The work of N.N.G.\ concerning the investigation of magnetic properties of samples was funded by the Russian Science
Foundation according to Research Project No.\ 21-72-20153-P.
The work of A.A.K., N.N.G., A.A.V., and I.A.G.\ concerning the investigation of transport properties of samples was financially supported by the government assignment for FRC Kazan Scientific Center of Russian Academy of Sciences.

\appendix*

\section{Optimal-fit parameter search method}
\label{AppendixPerturbation theory}

In this Appendix, we provide a detailed explanation of the fitting procedure.  
We start by defining the key parameters of our model \cite{Fominov2003}:  
\begin{equation}
    \xi_{S,F} \equiv \sqrt{\frac{\hbar D_{S,F}}{2\pi k_B T_{cS}}},\quad \gamma \equiv \frac{\rho_S \xi_S}{\rho_F \xi_F},\quad \gamma_b = \frac{R_b \mathcal{A}}{\rho_F \xi_F},
\end{equation}
where $\xi_{S,F}$, $D_{S,F}$, and $\rho_{S,F}$ are the coherence lengths, diffusion constants for conduction electrons, and the normal-state resistivities of the superconductor (ferromagnet), respectively. $T_{cS}$ is the superconducting transition temperature of an isolated S layer, and $\mathcal{A}$ is the interface area.
The key interface parameter $\gamma_b$ is proportional to the interface resistance $R_b$. The F layers are also characterized by the exchange energy $h$.

We now consider our method to find the values of the fitting parameters.
The primary challenge in applying the theoretical model to experimental
data lies in solving an inverse problem: while the model in Ref.\ \cite{Fominov2003}
determines $T_{c}$ based on known material parameters, the experiment
provides the value of $T_{c}$, requiring the corresponding material parameters
to be sought for. This search process presents two key difficulties.
First, since the interface thickness varies between samples within
each series, different values of the dimensionless resistance parameter
$\gamma_{b}$ are required for each sample and each series, significantly increasing
the number of fitting parameters. Second, if no quantitative fitting
is found within the entire parameter space, it remains unclear which
set of parameters should be preferred. The procedure described below,
which we employed to determine the fitted parameters and generate
theoretical plots, is designed to address these issues.

The fitting procedure begins with an estimation of the parameters' orders
of magnitude. The values of $\xi_{S}$, $\xi_{F}$, and $\gamma$
can be determined from the residual resistivity of the materials,
while the exchange energy is roughly estimated based on the Curie temperature
of iron. Since the materials are nominally the same within each series, the parameters $\xi_{S,F}$, $\gamma$, and $h$ are assumed to be identical for all samples across all series. Concerning $\gamma_{b}$, numerical simulations indicate that the experimentally
observed order of magnitude of the SSV effect is reproduced when $\gamma_{b}\sim 0.1$
for samples with intermediate insulating layer thicknesses. As a prerequisite condition, we impose the requirement that $\gamma_{b}$ must increase
with interface thickness within each series. 

The core idea of our approach is to systematically vary all material
parameters with a small step size within the estimated range, generating
a set of $T_{c}$ values corresponding to different parameter combinations.
The optimal set of parameters is then selected by minimizing the deviation
between the theoretical and experimental values of $T_{c}$, defined
as $\varepsilon=|T_{c}^\mathrm{expt}-T_{c}^\mathrm{theor}|.$ Since agreement must be
achieved for both the $T_{c}$ and $\Delta T_{c}$ curves, we simultaneously
minimize the weighted sum of errors for both plots.

The procedure can be described in more detail as follows. After generating
tables of $T_{c}$ values as functions of material parameters, we
determine the set of $\gamma_{b}$ values while keeping  $\xi_{S}$,  $\xi_{F}$, $\gamma$, and
$h$ fixed. For a given $d_\mathrm{Si_3 N_4}=0$, the value of
$\gamma_{b}$ is chosen to minimize the expression $a|T_{c}^\mathrm{expt}-T_{c}^\mathrm{theor}|+b|\Delta T_{c}^\mathrm{expt}-\Delta T_{c}^\mathrm{theor}|,$
where the weight coefficients $a$ and $b$ are manually adjusted.
For the next insulating layer thickness, the same procedure is applied,
but $\gamma_{b}$ is selected only among values greater than the one
obtained in the previous step. Once the set of $\gamma_{b}$ values
is determined for fixed material parameters, the errors are summed
over all interface thicknesses, and the resulting sum is minimized
with respect to $\xi_{S}$, $\xi_{F}$, $\gamma$, and $h$. The parameters
found through this optimization, which yield the smallest deviation,
are considered candidates for the final fit. This algorithm can be applied either within a single series or across
all the three series simultaneously. In the latter case, errors for the
same interface thickness are summed over the series. 

Candidate selection is performed by varying the error weights ($a$ and $b$) and
repeating the search procedure. Typically, for each data point within
a single series, minimization results in one of the two outcomes: either
$T_{c}$ quantitatively matches the experimental data but $\Delta T_{c}$
deviates by several orders of magnitude, or $\Delta T_{c}$ achieves
satisfactory quantitative agreement while $T_{c}$ matches the experimental
value by the order of magnitude. Among these possibilities, we always choose the second
option. 
It is important to note that this procedure, based on varying the error weights, results in only a finite, and relatively small, number of candidate parameter sets that provide an acceptable fit, even when all possible weight combinations are explored.

As a result, we have found three sets of optimal parameters differing in how the $\gamma_{b}$ values for the same interface thickness $d_\mathrm{Si_3 N_4}$ are correlated between different series. In
the first set, $\gamma_{b}$ was determined independently within each series (keeping our prerequisite condition that $\gamma_b$ should monotonically grow with increasing the interface thickness within each series). In the second set, an additional constraint was introduced: for
all $d_{2}>d_{1}$ and for all series indices $i,j$, the condition $\gamma_{b}^{(i)}(d_{2})>\gamma_{b}^{(j)}(d_{1})$
was enforced (monotonicity across all the series). In the third set, $\gamma_{b}(d_\mathrm{Si_3 N_4})$
was assumed to be independent of the series (strict correlation between the series).

\begin{figure}[t]
\center{\includegraphics[width=\columnwidth]{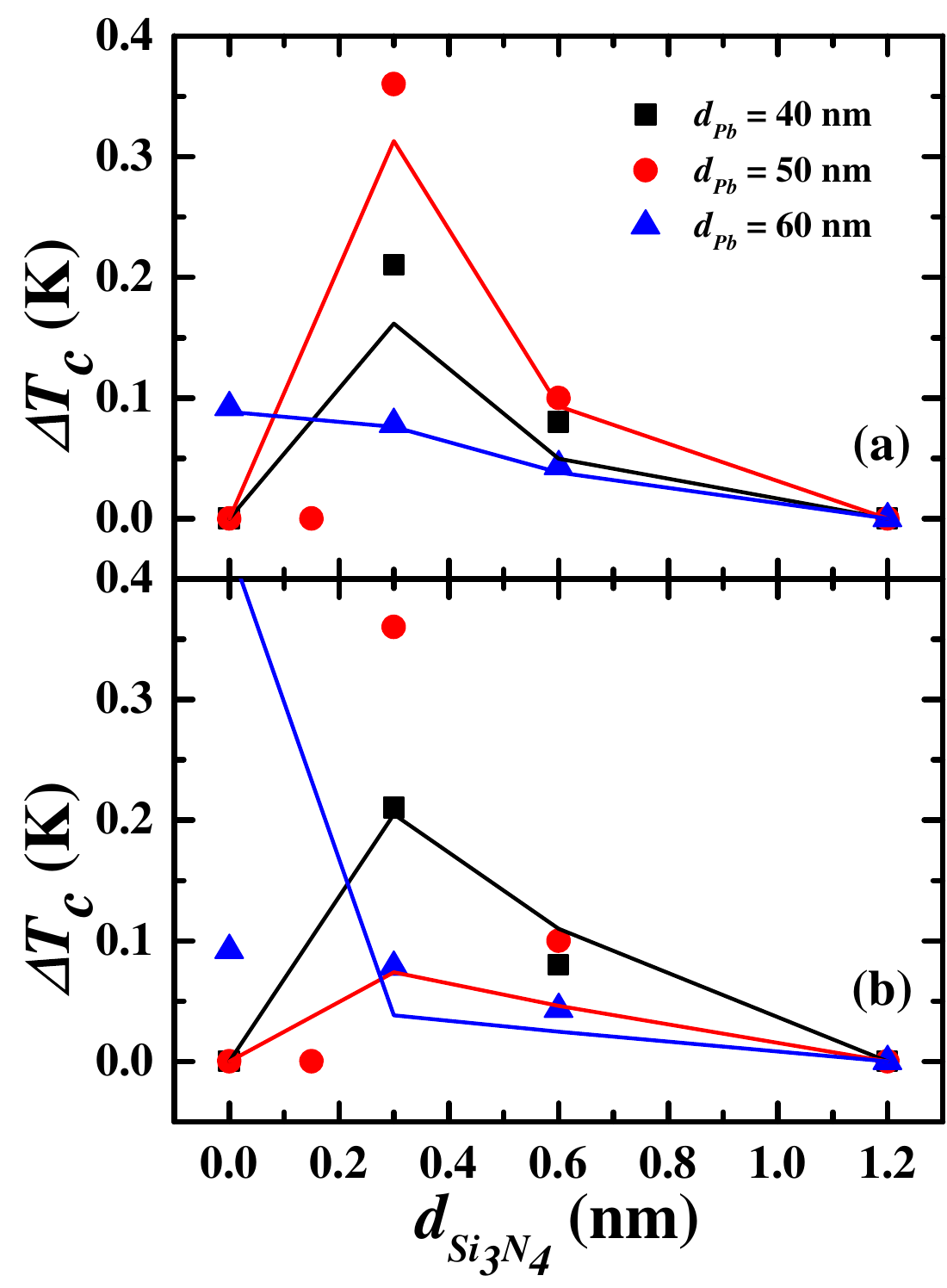}}
\caption{Illustration of different theoretical approaches to fitting $\Delta T_c$:
(a)~independent $\gamma_b(d_{\text{Si}_{3}\text{N}_{4}})$ dependencies within each series;
(b)~strict correlations between the series: $\gamma_b$ depends only on $d_{\text{Si}_{3}\text{N}_{4}}$ and not on the series.
The calculated theoretical points correspond only to $d_{\text{Si}_{3}\text{N}_{4}} =0$, $0.3$, $0.6$, and $1.2$\,nm; they are connected by straight solid lines to aid visual comparison with the experimental data. The fitting parameters are discussed in Appendix.}
\label{fig:DeltaTc_app_fit}
\end{figure}

\begin{table}[t] \centering
\begin{tabular}{|c|c|c|c|c|} 
\hline 
\textbf{$d_{\text{Si}_{3}\text{N}_{4}}=$} & \textbf{0\,\text{nm}} & \textbf{0.3\,\text{nm}} & \textbf{0.6\,\text{nm}} & \textbf{1.2\,\text{nm}} \\ 
\hline 
\textbf{Series 1} &0  &0.2  &0.29  &6.0  \\ 
\hline 
\textbf{Series 2} &0  &0.11  &0.18  &6.0  \\ 
\hline 
\textbf{Series 3} &0.14  &0.15  &0.2  &4.0  \\ 
\hline 
\end{tabular}
\caption{First set of optimal $\gamma_b$ values that results from independent fitting within each series. This set provides quantitative agreement of theoretical $\Delta T_c$ with experiment, see Fig.~\ref{fig:DeltaTc_app_fit}(a); however, $d_{\text{Si}_{3}\text{N}_{4}}$ is not always monotonic across the series of samples. The corresponding material parameters are $\xi_S = 40\,\text{nm}$, $\xi_F = 5\,\text{nm}$, $\gamma = 0.038$, and $h = 0.075\,\text{eV}$.} 
\label{GammaTab1fit} 
\end{table}

As the main result of our fitting procedure, we use the parameters of the second set (see Sec. \ref{Discussion} and Table \ref{GammaTab} in the main text). The first set, while providing quantitative agreement with experiment for $\Delta T_c$, see Fig. \ref{fig:DeltaTc_app_fit}(a), does not satisfy the condition of monotonicity across the series (see Table \ref{GammaTab1fit}), and was therefore rejected. With the third set, we qualitatively reproduce the $T_c$ and $\Delta T_c$ curves, see Fig. \ref{fig:DeltaTc_app_fit}(b). The corresponding fitted parameters are:  $\xi_S = 40\,\text{nm}$, $\xi_F = 5\:\text{nm}$, $\gamma = 0.05$, $h = 0.076\,\text{eV}$, while $\gamma_b$ for $d_\mathrm{Si_3 N_4}$ values of 0, 0.3, 0.6, and 1.2\,nm are 0.11, 0.26, 0.3, and 6, respectively. However, in Series~3, $\Delta T_c(d_\mathrm{Si_3 N_4} = 0)$ significantly deviates from the experimental data, and in Series~2, the values of $\Delta T_c$ become even lower than those resulting from the second set. At the same time, taking into account that the interface formation process does not guarantee a precise number of atomic layers and their uniformity, we regard fluctuations of $\gamma_b$ across the series (at the same nominal interface thickness) as acceptable, provided that monotonicity is preserved. Therefore, the final fitted parameters used in the main text are taken from the second set.

The entire procedure described above employed the multimode method for calculating $T_c$ for a given set of material parameters, as detailed in Refs.\ \cite{Fominov2002,Fominov2001}. Generally speaking, the multimode search for $T_{c}$ reduces to finding
a point where the determinant of a large matrix vanishes. While direct
computation is feasible, it is computationally expensive. To efficiently
narrow down the parameter space, we first performed the entire algorithm
within the single-mode approximation. 

Over a broad parameter range, the single-mode approximation differs
from the multimode result by only a few percent, yet it accelerates
the computation of $T_{c}$ by factors of tens to hundreds, depending
on the number of modes used. In our case, we employed 30 modes. Once
an approximate parameter set was obtained using the single-mode method,
we refined the search in its vicinity within the full multimode approach.


\begin{thebibliography}{99}

\bibitem{Oh}
S. Oh, D. Youm, and M.~R. Beasley,
{A superconductive magnetoresistive memory element using controlled exchange interaction},
\href{https://doi.org/10.1063/1.120032}{Appl. Phys. Lett. \textbf {71}, {2376} (1997).}

\bibitem{Tagirov}
L.~R. Tagirov,
Low-field superconducting spin switch based on a superconductor/ferromagnet multilayer,
\href {https://doi.org/10.1103/PhysRevLett.83.2058} {{Phys. Rev. Lett.} \textbf {83}, {2058} (1999).}

\bibitem{Buzdin2}
A.~I. Buzdin, A.~V. Vedyayev, and N.~V. Ryzhanova,
{Spin-orientation–dependent superconductivity in F/S/F structures},
\href {http://stacks.iop.org/0295-5075/48/i=6/a=686} {{Europhys. Lett.} \textbf {48}, {686} (1999).}

\bibitem{Ioffe}  
L.~B. Ioffe, V.~B. Geshkenbein, M.~V. Feigel'man, A.~L. Fauchère, and G. Blatter,
{Environmentally decoupled sds-wave Josephson junctions for quantum computing},
\href{https://doi.org/10.1038/19464} {Nature (London) \textbf{398}, {679} (1999).}

\bibitem{Feigelman}
M.~V. Feigel'man,
{A quantum bit based  on a Josephson contact between conventional and high-temperature
 superconductors (theory)},
\href{https://doi.org/10.1070/PU1999v042n08ABEH000599} {{Phys. Usp.} \textbf {42}, {823} (1999).}

\bibitem{Buzdin1} 
A.~I. Buzdin,
{Proximity effects in superconductor-ferromagnet heterostructures},
\href{https://doi.org/10.1103/RevModPhys.77.935} {{Rev. Mod. Phys.} \textbf {77}, {935} (2005).}

\bibitem{Linder}
J.~Linder and J.~W.~A. Robinson, 
{Superconducting spintronics},
\href {https://doi.org/10.1038/nphys3242} {Nat. Phys. \textbf {11}, {307} (2015)}.

\bibitem{Eschrig}
M. Eschrig, 
{Spin-polarized supercurrents for spintronics: a review of current progress},
\href {https://doi.org/10.1088/0034-4885/78/10/104501} {{Rep. Prog. Phys.} \textbf {78}, {104501} (2015).}

\bibitem{Ryazanov1999}
V.~V. Ryazanov,
{Josephson superconductor-ferromagnet-superconductor $\pi$-contact as an element of a quantum bit (experiment)},
\href {https://doi.org/10.1070/PU1999v042n08ABEH000600} {{Physics-Uspekhi} \textbf {42}, {825} (1999).}

\bibitem{Ryazanov2001a} 
V.~V. Ryazanov, V.~A. Oboznov, A.~Yu. Rusanov, A.~V. Veretennikov, A.~A. Golubov, and J. Aarts,
{Coupling of Two Superconductors through a Ferromagnet: Evidence for a $\pi$ Junction},
\href {https://doi.org/10.1103/PhysRevLett.86.2427} {{Phys. Rev. Lett.} \textbf{86}, {2427} (2001).}

\bibitem{Gu2002}
J.~Y. Gu, C.-Y. You, J.~S. Jiang, J. Pearson, Ya.~B. Bazaliy, and S.~D. Bader,
Magnetization-Orientation Dependence of the Superconducting Transition Temperature in the
Ferromagnet-Superconductor-Ferromagnet System: CuNi/Nb/CuNi,
\href{https://doi.org/10.1103/PhysRevLett.89.267001}
{Phys. Rev. Lett. \textbf{89}, 267001 (2002)}.

\bibitem{Zdravkov2013}
V.~I. Zdravkov, D. Lenk, R. Morari, A. Ullrich, G. Obermeier, C. M\"uller, H.-A.~Krug von Nidda, A.~S. Sidorenko, S. Horn, R. Tidecks, and L.~R. Tagirov,
{Memory effect and triplet pairing generation in the superconducting exchange biased Co/\allowbreak CoO$_x$/\allowbreak Cu$_{41}$Ni$_{59}$/\allowbreak Nb/\allowbreak Cu$_{41}$Ni$_{59}$ layered heterostructure},
\href {https://doi.org/10.1063/1.4818266} {{Appl. Phys. Lett.} \textbf {103}, {062604} (2013).}

\bibitem{Lenk2017}
D. Lenk, R. Morari, V.~I. Zdravkov, A. Ullrich, Yu. Khaydukov, G. Obermeier, C. M\"uller, A.~S. Sidorenko, H.-A. Krug von Nidda, S. Horn, L.~R. Tagirov, and R. Tidecks,
{Full-switching FSF-type superconducting spin-triplet magnetic random access memory element},
\href {https://doi.org/10.1103/PhysRevB.96.184521} {{Phys. Rev. B} \textbf {96}, {184521} (2017).}

\bibitem{Gaifullin} 
R.~R. Gaifullin, R.~G. Deminov, M.~N. Aliyev, and L.~R. Tagirov,
{Superconducting spin-valves in spintronics},
\href {https://doi.org/10.26907/mrsej-19304} {{Magn. Reson. Solids} \textbf {21}, {19304} (2019).}

\bibitem{Kamashev20191}
A.~A. Kamashev, N.~N. Garif'yanov, A.~A. Validov, J. Schumann, V. Kataev, B. B\"{u}chner, Ya.~V. Fominov, and I.~A. Garifullin,
{Superconducting spin-valve effect in heterostructures with ferromagnetic Heusler alloy layers}, 
\href {https://doi.org/10.1103/PhysRevB.100.134511} {{Phys. Rev. B} \textbf {100}, {134511} (2019).}

\bibitem{Kamashev20241}
A.~A. Kamashev, N.~N. Garif'yanov, A.~A. Validov, V. Kataev, A.~S. Osin, Ya.~V. Fominov,
and I.~A. Garifullin,
{Expanding the operational temperature window of a superconducting spin valve}, 
\href {https://doi.org/10.1103/PhysRevB.109.144517} {{Phys. Rev. B} \textbf {109}, {144517} (2024).}

\bibitem{Eschrig2} M.~Eschrig,
{Spin-polarized supercurrents for spintronics}, 
\href {https://doi.org/10.1063/1.3541944} {{Phys. Today} \textbf {64}, {43} ({2011}).}

\bibitem{Moraru}
I.~C. Moraru, W.~P. Pratt, and N.~O. Birge,
{Magnetization-Dependent $T_c$ Shift in Ferromagnet/Superconductor/Ferromagnet Trilayers with a Strong Ferromagnet},
\href {https://doi.org/10.1103/PhysRevLett.96.037004} {Phys. Rev. Lett. \textbf{96}, 037004 (2006).}

\bibitem{Steiner} 
R. Steiner and P. Ziemann,
{Magnetic switching of the superconducting transition temperature in layered ferromagnetic/superconducting hybrids: Spin switch versus stray field effects},
\href {https://doi.org/10.1103/PhysRevB.74.094504} {{Phys. Rev. B} \textbf{74}, {094504} (2006).}

\bibitem{Stoutimore}
M.~J.~A. Stoutimore, A.~N. Rossolenko, V.~V. Bolginov, V.~A. Oboznov, A.~Y. Rusanov, D.~S. Baranov, N.~G. Pugach, 
S.~M. Frolov, V.~V. Ryazanov, and D.~J. Van~Harlingen,
{Second-Harmonic Current-Phase Relation in Josephson Junctions with Ferromagnetic Barriers},
\href {https://doi.org/10.1103/PhysRevLett.121.177702} {{Phys. Rev. Lett.} \textbf {121}, {177702} (2018).}

\bibitem{Pugach2009}
N.~G. Pugach, M.~Yu. Kupriyanov, A.~V. Vedyayev, C. Lacroix, E. Goldobin, D. Koelle, R. Kleiner, and A.~S. Sidorenko,
{Ferromagnetic Josephson junctions with steplike interface transparency},
\href {https://doi.org/10.1103/PhysRevB.80.134516} {{Phys. Rev. B} \textbf {80}, {134516} (2009).}

\bibitem{Leksin2010}
P.~V. Leksin, N.~N. Garif’yanov, I.~A. Garifullin, J. Schumann, H. Vinzelberg, V. Kataev, R. Klingeler,
O.~G. Schmidt, and B. B\"{u}chner,
{Full spin switch effect for the superconducting current in a superconductor/ferromagnet thin film heterostructure},
\href {https://doi.org/10.1063/1.3486687} {{Appl. Phys. Lett.} \textbf {97}, {102505} ({2010}).}

\bibitem{Fominov} 
Ya.~V. Fominov, A.~A. Golubov, T.~Yu. Karminskaya, M.~Yu. Kupriyanov, R.~G. Deminov, and L.~R. Tagirov,
{Superconducting triplet spin valve},
\href {https://doi.org/10.1134/S002136401006010X} {{JETP Lett.} \textbf {91}, {308} (2010).}

\bibitem{Leksin2011}
P.~V. Leksin, N.~N. Garif’yanov, I.~A. Garifullin, J. Schumann, V. Kataev, O.~G. Schmidt, and B. B\"{u}chner,
{Manifestation of new interference effects in a superconductor-ferromagnet spin valve},
\href {https://doi.org/10.1103/PhysRevLett.106.067005} {{Phys. Rev. Lett.} \textbf {106}, {067005} ({2011}).}

\bibitem{Kamashev2019}
A.~A. Kamashev, N.~N. Garif'yanov, A.~A. Validov, J. Schumann, V. Kataev, B.~B\"{u}chner, Ya.~V. Fominov, and I.~A. Garifullin,
{Superconducting switching due to a triplet component in the Pb/\allowbreak Cu/\allowbreak Ni/\allowbreak Cu/\allowbreak Co$_2$Cr$_{1-x}$Fe$_x$Al$_y$ spin-valve structure},
\href {https://doi.org/10.3762/bjnano.10.144} {{Beilstein J. Nanotechnol.} \textbf {10}, {1458} (2019).}

\bibitem{Gu2015}
Y. Gu, G.~B. Halasz, J.~W.~A. Robinson, and M.~G. Blamire,
{Large superconducting spin valve effect and ultrasmall exchange splitting in epitaxial rare-earth-niobium trilayers},
\href {https://doi.org/10.1103/PhysRevLett.115.067201} {{Phys. Rev. Lett.} \textbf{115}, {067201} ({2015}).}

\bibitem{Aarts2015}
A. Singh, S. Voltan, K. Lahabi, and J. Aarts,
{Colossal Proximity Effect in a Superconducting Triplet Spin Valve Based on the Half-Metallic Ferromagnet CrO$_2$},
\href {https://doi.org/10.1103/PhysRevX.5.021019} {{Phys. Rev. X} \textbf {5}, {021019} ({2015}).}

\bibitem{Nano}
P.~V. Leksin, A.~A. Kamashev, J. Schumann, V.~E. Kataev, J. Thomas, B.~B\"{u}chner, and I.~A. Garifullin, 
{Boosting the superconducting spin valve effect in a metallic superconductor/ferromagnet heterostructure},
\href {https://doi.org/10.1007/s12274-016-0988-y} {{Nano Res.} \textbf {9},  {1005} ({2016}).}

\bibitem{Halterman4}
K. Halterman and M. Alidoust, 
{Half-metallic superconducting triplet spin valve},
\href {https://doi.org/10.1103/PhysRevB.94.064503} {{Phys. Rev. B} \textbf {94}, {064503} ({2016}).}

\bibitem{Halterman5}
M. Alidoust, K. Halterman, and O.~T. Valls,
{Zero-energy peak and triplet correlations in nanoscale superconductor/ferromagnet/ferromagnet spin valves},
\href {https://doi.org/10.1103/PhysRevB.92.014508} {{Phys. Rev. B} \textbf {92}, {014508} ({2015}).}

\bibitem{DeGennes1966}
P.G. {De Gennes},
{Coupling between ferromagnets through a superconducting layer},
\href {https://doi.org/10.1016/0031-9163(66)90229-0} {{Phys. Lett.} \textbf {23}, {10} (1966).} 

\bibitem{Li2013} 
B. Li, N. Roschewsky, B.~A. Assaf, M. Eich, M. Epstein-Martin, D. Heiman, M. Münzenberg, and J.~S. Moodera, 
{Superconducting Spin Switch with Infinite Magnetoresistance Induced by an Internal Exchange Field},
\href {https://doi.org/10.1103/PhysRevLett.110.097001} {{Phys. Rev. Lett.} \textbf {110}, {097001} ({2013}).}

\bibitem{Zhu2017}
Y. Zhu, A. Pal, M.~G. Blamire, and Z.~H. Barber,
{Superconducting exchange coupling between ferromagnets},
\href {https://doi.org/10.1038/nmat4753} {{Nat. Mater.} \textbf {16}, {195} (2017).} 

\bibitem{Deutscher} 
G. Deutscher and F. Meunier,
{Coupling Between Ferromagnetic Layers Through a Superconductor},
\href {https://doi.org/10.1103/PhysRevLett.22.395} {{Phys. Rev. Lett.} \textbf {22}, {395} ({1969}).}

\bibitem{Miao2007}
G.-X. Miao, K. Yoon, T.~S. Santos, and J.~S. Moodera,
{Influence of Spin-Polarized Current on Superconductivity and the Realization of Large Magnetoresistance},
\href {https://doi.org/10.1103/PhysRevLett.98.267001} {{Phys. Rev. Lett.} \textbf {98}, {267001} (2007).} 

\bibitem{Luo2010}
Y. Luo and K. Samwer,
{Superconductive spin-valve effect in CoFeHf/Pb/CoFeHf layered structures},
\href {https://doi.org/10.1209/0295-5075/91/37003} {{Europhys. Lett.} \textbf {91}, {37003} (2010).} 

\bibitem{Kamashev20242}
A.~A. Kamashev, N.~N. Garif'yanov, A.~A. Validov, V. Kataev, A.~S. Osin, Ya.~V. Fominov, and I.~A. Garifullin,
{Superconducting spin valve effect in a Co/Pb/Co heterostructure with insulating spacers},
\href {https://doi.org/10.1134/S0021364024600095} {{JETP Lett.} \textbf {119}, {305} (2024).}

\bibitem{Kamashev2024}
A.~A. Kamashev, N.~N. Garif'yanov, A.~A. Validov, V. Kataev, A.~S. Osin, Ya.~V. Fominov, and I.~A. Garifullin,
{Superconducting spin valve effect in Co/Pb/Co heterostructures with insulating interlayers},
\href {https://doi.org/10.3762/bjnano.15.41} {{Beilstein J. Nanotechnol.} \textbf {15}, {457} (2024).}

\bibitem{Baladie2003}
I. Baladi\'e and A. Buzdin,
Thermodynamic properties of ferromagnet/superconductor/ferromagnet nanostructures, 
\href{https://doi.org/10.1103/PhysRevB.67.014523}
{Phys. Rev. B \textbf{67}, 014523 (2003)}.

\bibitem{Fominov2003}
Ya.~V. Fominov, A.~A. Golubov, and M.~Yu. Kupriyanov, 
{Triplet proximity effect in FSF trilayers},
\href {https://doi.org/10.1134/1.1591981} {{JETP Lett.} \textbf {77}, {510} (2003).}

\bibitem{Fominov2002}
Ya.~V. Fominov, N.~M. Chtchelkatchev, and A.~A. Golubov,
{Nonmonotonic critical temperature in superconductor/ferromagnet bilayers},
\href {http://dx.doi.org/10.1103/PhysRevB.66.014507} {{Phys. Rev. B} \textbf {66}, {014507} (2002).}

\bibitem{Fominov2001}
Ya.~V. Fominov, N.~M. Chtchelkatchev, and A.~A. Golubov,
{Critical temperature of superconductor/ferromagnet bilayers},
\href {https://doi.org/10.1134/1.1405893} {JETP Lett. \textbf{74}, 96 (2001).}

\end{thebibliography}
\end{document}